\documentclass[aps,prd,10pt,twocolumn,preprintnumbers,superscriptaddress,bibnotes,nofootinbib,letterpaper,longbibliography]{revtex4-2}

\usepackage{hyperref}
\usepackage{graphicx}
\usepackage{amsfonts,amsmath,amssymb,bm}
\usepackage{physics}
\usepackage{svg}
\usepackage{array}
\usepackage[normalem]{ulem}
\usepackage[table]{xcolor} 

\hypersetup{
  colorlinks  = true,
  urlcolor    = blue,
  linkcolor   = blue,
  citecolor   = red
}

\newcommand{\orcid}[1]
{\begingroup
  \hypersetup{hidelinks}\href{https://orcid.org/#1}{\includegraphics[width=9pt]{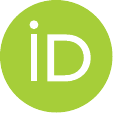}
} \endgroup}

\usepackage{booktabs}

\begin{document}

\title{New Spallation Background Rejection Techniques to \\ Greatly Improve the Solar Neutrino Sensitivity of JUNO}

\author{Obada Nairat \orcid{0000-0003-2019-9021}}
\email{nairat.2@osu.edu}
\affiliation{Center for Cosmology and AstroParticle Physics (CCAPP), \href{https://ror.org/00rs6vg23}{Ohio State University}, Columbus, OH 43210}
\affiliation{Department of Physics, \href{https://ror.org/00rs6vg23}{Ohio State University}, Columbus, OH 43210}

\author{John F. Beacom \orcid{0000-0002-0005-2631}}
\email{beacom.7@osu.edu}
\affiliation{Center for Cosmology and AstroParticle Physics (CCAPP), \href{https://ror.org/00rs6vg23}{Ohio State University}, Columbus, OH 43210}
\affiliation{Department of Physics, \href{https://ror.org/00rs6vg23}{Ohio State University}, Columbus, OH 43210}
\affiliation{Department of Astronomy, \href{https://ror.org/00rs6vg23}{Ohio State University}, Columbus, OH 43210} 

\author{Shirley Weishi Li \orcid{0000-0002-2157-8982}}
\email{shirley.li@uci.edu}
\affiliation{Department of Physics and Astronomy, \href{https://ror.org/04gyf1771}{University of California}, Irvine, CA 92697}

\date{October 14, 2025}

\begin{abstract}
While the potential of the Jiangmen Underground Neutrino Observatory (JUNO) to measure solar neutrinos is known, realizing this potential requires new techniques to reduce detector backgrounds.  One of the most serious backgrounds is due to the beta decays of unstable nuclei produced through muon breakup (spallation) of nuclei.  This background is much more significant in JUNO compared to Super-Kamiokande due to JUNO's shallower depth and its lack of directional information.  We present the first detailed theoretical calculations of spallation backgrounds in JUNO, showing the underlying physical processes and new ways to cut backgrounds while preserving signals.  A key point is showing the importance of neutron tagging to identify hadronic showers, which are rare but produce almost all of the dangerous isotopes.  \textit{With our new techniques, JUNO will be able to reduce deadtime (signal loss) by a factor of about 4.5 and to reduce the running time needed to meet sensitivity goals by a factor of two.}  This will give JUNO greatly improved sensitivity to $^8$B and $hep$ solar neutrinos, as we will explore in a separate paper.
\end{abstract}

\maketitle

%%%%%%%%%%%%%%%%%%%%%%%%%%%%%%%%%%%%%%%%%%%%%%%%%%%%%%%%%%%%%%%%%%%%%%%
%%%%%%%%%%%%%%%%%%%%%%%%%%%%%%%%%%%%%%%%%%%%%%%%%%%%%%%%%%%%%%%%%%%%%%%

\section{Introduction}
\label{sec:Introduction}

Studies of neutrino mixing are entering a precision era, driven by the capabilities of huge new detectors.  Starting in 2025, the Jiangmen Underground Neutrino Observatory (JUNO; inner detector mass of 20~kton) will use reactor antineutrinos to probe $\sin^2\theta_{12}$, $\Delta m_{21}^2$, and $\Delta m_{31}^2$, the latter including the neutrino mass ordering~\cite{JUNO:2015zny, JUNO:2022mxj}.  Then, starting in 2028, Hyper-Kamiokande (Hyper-K; 216~kton) will use accelerator neutrinos to probe $\sin^2\theta_{23}$, $\Delta m_{32}^2$, and $\delta_\text{CP}$~\cite{Fukasawa:2016yue, Hyper-Kamiokande:2018ofw}.  Finally, around 2030, the first two modules of the Deep Underground Neutrino Experiment (DUNE; 35~kton) will do the same with higher statistics and unique event-reconstruction capabilities~\cite{DUNE:2020lwj, DUNE:2020jqi}.  For comparison, Super-Kamiokande (Super-K), which has long been the only huge precision neutrino detector, has an inner detector mass of 32~kton~\cite{Super-Kamiokande:2002weg, Abe:2013gga}.

Importantly, these new detectors can also make precise measurements of neutrino mixing parameters using solar neutrinos~\cite{Hyper-Kamiokande:2018ofw, Capozzi:2018dat, JUNO:2020hqc, JUNO:2022jkf, JUNO:2023zty, Parsa:2022mnj, Meighen-Berger:2024xbx}.  \textit{Comparing mixing parameters from laboratory versus astrophysical sources will allow unprecedented tests of new physics.}  Here we focus on ``high-energy" solar neutrinos, meaning the $^8$B and $hep$ fluxes.  Even after outstanding work by previous experiments, especially Super-K and the Sudbury Neutrino Observatory (SNO)~\cite{Super-Kamiokande:2005wtt, SNO:2011hxd, SNO:2020gqd, Super-Kamiokande:2023jbt}, key questions remain.  Answering these will better probe the mixing parameters.
\begin{enumerate}

    \item The expected energy dependence of the electron neutrino survival probability has not been convincingly seen, despite this being one of the most important predictions of the Mikheyev-Smirnov-Wolfenstein (MSW) theory~\cite{Wolfenstein:1977ue, Mikheyev:1985zog}.  The data at low and high energies are separately consistent with being flat, with inadequate data in the transition region~\cite{BOREXINO:2018ohr}. Super-K sees hints of the expected rise in the survival probability towards decreasing energies~\cite{Super-Kamiokande:2023jbt}, but SNO sees hints of a drop~\cite{SNO:2011hxd}.

    \item The day-night asymmetry --- another key prediction of the MSW theory --- has been detected by Super-Kamiokande at 2.4$\sigma$~\cite{Super-Kamiokande:2023jbt}, so greater precision is needed.  This and the measurement above are critical to determining $\Delta m_{21}^2$, which presently has a 1.5-$\sigma$ discrepancy between reactor and solar measurements, with the reactor value being higher.  Even if this discrepancy shrinks, testing new physics depends on improving the solar precision to match the huge improvements in the reactor precision expected from JUNO.

    \item While both Super-K and SNO have set competitive limits on the $hep$ flux~\cite{Super-Kamiokande:2005wtt, SNO:2020gqd}, it has yet to be detected.  It is important because it originates from larger radii in the Sun than the $^8$B flux, leading to significant changes in the survival probability at the same energy~\cite{Denton:2025jkt, Zaidel:2025kdk}.  Because the $hep$ flux is $\sim$$10^{-3}$ times the $^8$B flux, it can be isolated only via its higher endpoint energy.
    
\end{enumerate}

While all of these new detectors (plus Super-K, for which the recent addition of dissolved gadolinium gives it new sensitivity~\cite{Beacom:2003nk, Super-Kamiokande:2024kcb}) are expected to make major progress on the three questions above, JUNO has a shorter timeline and special capabilities.  Because JUNO is scintillator-based, its light yield is about 100 times greater than Super-K's, leading to better energy resolution and the possibility of a lower energy threshold.  While Borexino and KamLAND were also scintillator-based, their inner detector masses were much smaller (0.3~kton and 1~kton).  The huge volume of JUNO helps not only with exposure, but also with allowing thicker shielding from external radioactivity.

To address the first question above, it is necessary to observe the $^8$B spectrum to lower energies and with lower backgrounds.  To address the second, greater exposure and lower backgrounds are needed.  To address the third, greater exposure and better energy resolution are needed.  JUNO is well-positioned to attack all three questions.  However, this will only be possible with extensive new work on background reduction, because JUNO has two key disadvantages.  Compared to Super-K (at a depth of 2700 meters water equivalent), JUNO is at a shallower depth (1800 m.w.e.), which means that muon-induced radioactivity (due to ``spallation," or nuclear breakup processes) is a more serious problem.  Second, unlike Super-K, JUNO cannot cut backgrounds based on their direction relative to the Sun, because scintillators isotropize the light from neutrino-electron scattering. While recent reconstruction techniques have successfully demonstrated event-by-event directional reconstruction in liquid scintillators~\cite{BOREXINO:2021efb, SNO:2023cnz}, and future `hybrid' detectors like THEIA~\cite{Theia:2019non} propose to further exploit Cherenkov-scintillation separation, these techniques are not yet the baseline for JUNO.

In this paper, we develop new background-reduction techniques to help JUNO to realize its discovery potential for solar neutrinos, focusing on the energy range above 2.3~MeV (this choice is explained in Secs.~\ref{sec:Muons} and~\ref{sec:Isotopes}).  We present the most detailed theoretical calculations of spallation for liquid-scintillator detectors, building on similar calculations by our group for water-based detectors like Super-K, with carryover to Hyper-Kamiokande~\cite{Li:2014sea, Li:2015kpa, Li:2015lxa, Nairat:2024upg}, plus for liquid-argon-based detectors like DUNE~\cite{Zhu:2018rwc}.  By elucidating the underlying physical processes, our work here shows how to develop cuts that significantly improve on those developed empirically from experimental data alone.  Our goals are to show how to increase cut efficiency (the fraction of background removed) and decrease cut deadtime (the fraction of signal removed).  This work will thus be critical to JUNO's ability to address the three physics questions above. 

The key new ingredient in our work is using neutron tagging to identify hadronic showers, which are rare but produce the vast majority of dangerous isotopes.  As we explain below (Sec.~\ref{sec:Showers}), our proposed new technique is substantially different from a standard technique (TFC, or a three-fold coincidence between muon, neutron, and candidate isotope) used to reject $^{11}$C and $^{10}$C in scintillator detectors~\cite{Galbiati:2004wx, Borexino:2011cjz, Borexino:2021pyz, KamLAND-Zen:2023spw}.  \textit{Our technique is much more general, focusing on hadronic showers instead of just the specifics of isotope production.}  The physical insights we develop for JUNO carry over to other scintillator detectors like Borexino~\cite{Borexino:2013cke, Borexino:2017uhp}, KamLAND~\cite{KamLAND:2009zwo, KamLAND:2011fld}, and SNO+~\cite{SNO:2015wyx,SNO:2024vjl}, though Borexino and SNO+ have less of a problem with spallation backgrounds due to their greater depths.

The rest of this paper is organized as follows. In Sec.~\ref{section:Review}, we briefly review solar neutrino signals and backgrounds. In Sec.~\ref{sec:Muons}, we present the setup used to simulate muon propagation in JUNO. In Sec.~\ref{sec:Isotopes}, we calculate the yields and spectra of spallation isotopes. In Sec.~\ref{sec:Showers}, we present detailed descriptions of shower and neutron production and their relations to isotope production. In Sec.~\ref{sec:Cuts}, we propose some improved spallation cuts. Finally, we conclude in Sec.~\ref{sec:Conclusions}.

In a separate paper, we will discuss the implications of this work in detail and calculate its impact on JUNO's sensitivity to the transition region of the electron neutrino survival probability, day-night asymmetry, and its potential to detect $hep$ solar neutrinos.

%%%%%%%%%%%%%%%%%%%%%%%%%%%%%%%%%%%%%%%%%%%%%%%%%%%%%%%%%%%%%%%%%%%%%%%
%%%%%%%%%%%%%%%%%%%%%%%%%%%%%%%%%%%%%%%%%%%%%%%%%%%%%%%%%%%%%%%%%%%%%%%

\section{Review of High-Energy Solar Neutrino Signals and Backgrounds}
\label{section:Review}

Because the MSW effects increase with energy, the $^8$B (endpoint $\sim$15~MeV) and $hep$ (endpoint $\sim$19~MeV) neutrino fluxes are especially powerful probes of neutrino mixing in the Sun and Earth~\cite{Wolfenstein:1977ue, Mikheyev:1985zog}.  The $^8$B neutrinos are produced through beta decays of those nuclei, primarily in the solar radial range 0--0.1~$R_\odot$, while the $hep$ neutrinos are produced through the weak proton capture reaction on $^3$He, primarily in the solar radial range 0--0.3~$R_\odot$~\cite{Bahcall:2004pz, Vinyoles:2016djt, Acharya:2024lke}.  

The most precise measurements of the $^8$B flux --- and the most stringent limits on the $hep$ flux --- have been made by Super-K and SNO~\cite{Super-Kamiokande:2023jbt, SNO:2011hxd, SNO:2020gqd, Super-Kamiokande:2005wtt}.  Super-K is a water-based experiment with an inner detector mass of 32~kton~\cite{Super-Kamiokande:2002weg, Abe:2013gga}. It detects solar neutrinos via neutrino-electron scattering through the Cherenkov light emitted by recoil electrons, enabling precise directional reconstruction. SNO was a 1~kton detector with heavy water as a target material, which offered more detection channels through charged and neutral current interactions on deuterons, including enabling flavor-independent measurements~\cite{SNO:1999crp, Bellerive:2016byv}.  In both detectors, the detector backgrounds generally fall with increasing energy.  Accordingly, both are limited in how low in energy they have probed the $^8$B flux; both reached 3.5~MeV~\cite{SNO:2009uok, Super-Kamiokande:2023jbt}.  (For comparison, Borexino reached 3~MeV but with much lower statistics~\cite{Borexino:2008fkj}.)  At lower energies, there are overwhelming backgrounds from the beta decays of intrinsic radioactivity.

In the following, we focus on spallation radioactivity backgrounds, for which new theoretical work can lead to lower backgrounds and better sensitivity.  Despite being located at a great depth of 2700~m.w.e., Super-K experiences a cosmic-ray muon rate of approximately 2~s$^{-1}$~\cite{Super-Kamiokande:2023jbt}. These muons occasionally interact with oxygen nuclei in water, producing long-lived radioactive isotopes whose beta decays mimic signal events.  In Super-K, spallation-induced radioactivity is the dominant background (after cuts) above about 6~MeV~\cite{Super-Kamiokande:2015xra, Locke:2020kco, Coffani:2021gbe, Super-Kamiokande:2021snn}.  (In SNO, they are almost negligible due to the greater overburden.)  The physical insights obtained in previous theoretical work on Super-K~\cite{Li:2014sea, Li:2015kpa, Li:2015lxa, Nairat:2024upg} have great relevance for understanding spallation backgrounds in JUNO.  

To suppress spallation backgrounds, Super-K utilizes the correlations between MeV-range candidate events and preceding GeV-range muon events~\cite{Bays:2012wty, Super-Kamiokande:2015xra, Coffani:2021gbe, Super-Kamiokande:2021snn, Locke:2020kco}.  The simplest approach is to define a virtual cylindrical veto region surrounding muon tracks that enter the detector, within which all subsequent candidate events are rejected for a certain time window. The size of the cylinder and the duration of the veto are chosen based on the typical spatial displacements and lifetimes of spallation products. Larger cylindrical cuts with longer time windows provide greater cut efficiency but also greater deadtime.  In practice, Super-K uses a sophisticated likelihood-based approach instead of simple cylinder cuts~\cite{Bays:2012wty, Coffani:2021gbe, Super-Kamiokande:2021snn, Locke:2020kco}.  A major strength of the Super-K detector is its ability to distinguish solar neutrino events from isotropic backgrounds using the angular correlation between the scattered electron and the direction of the Sun. However, even in the forward direction, the signal-to-background ratio after cuts remains close to unity, which poses significant limitations on sensitivity~\cite{Super-Kamiokande:2023jbt}.

Nearly all spallation isotopes are not produced directly by the muons themselves, but rather in secondary hadronic and electromagnetic showers initiated by muon radiative losses~\cite{Li:2014sea}. Importantly, the probability of isotope production increases with the energy of these showers~\cite{Li:2015kpa, Locke:2020kco, Coffani:2021gbe}. This understanding enables a more targeted approach: instead of applying vetoes after all muons, cuts can be restricted to muons associated with high-energy showers, significantly reducing the resulting deadtime. Moreover, the Cherenkov light profile along the muon track can be analyzed to localize the spatial position of these showers~\cite{Bays:2012wty, Li:2015kpa, Li:2015lxa, Super-Kamiokande:2021snn, Locke:2020kco, Coffani:2021gbe}, allowing for smaller veto regions. These strategies together reduce the overall impact on exposure while maintaining strong background rejection.  In Super-K's present likelihood-based cuts, the probability of a candidate event being due to spallation is evaluated using parameters such as the preceding muon's energy loss, plus the event's time delay, transverse displacement from the muon's track, and longitudinal displacement from the muon's light profile peak~\cite{Super-Kamiokande:2021snn, Super-Kamiokande:2023jbt}.

In recent work, we demonstrated that neutron tagging provides a powerful new approach to reducing the impact of spallation backgrounds~\cite{Nairat:2024upg}. \textit{Because many spallation isotopes are produced in hadronic showers along with neutrons, tagging neutron captures is a powerful way to reduce backgrounds.  Because hadronic showers are rare, focusing on them can also reduce deadtime.}  This promises future improvements in low-energy solar neutrino sensitivity in Super-K with gadolinium.  JUNO and other scintillator detectors also have neutron tagging ability, allowing similar cuts.

%%%%%%%%%%%%%%%%%%%%%%%%%%%%%%%%%%%%%%%%%%%%%%%%%%%%%%%%%%%%%%%%%%%%%%%
%%%%%%%%%%%%%%%%%%%%%%%%%%%%%%%%%%%%%%%%%%%%%%%%%%%%%%%%%%%%%%%%%%%%%%%

\section{Muon irradiation of JUNO}
\label{sec:Muons}

In this section, we describe the JUNO detector and the simulation setup we use to study its exposure to cosmic-ray muons and their energy depositions.

%%%%%%%%%%%%%%%%%%%%%%%%%%%%%%%%%%%%%%%%%%%%%%%%%%%%%%%%%%%%%%%%%%%%%%%

\subsection{JUNO design and backgrounds}

JUNO is a multi-purpose liquid scintillator neutrino detector located 53~km away from both the Yangjiang and Taishan Nuclear Power Plants in Guangdong, China, shielded under $\sim$650~m of overburden ($\sim$1800 m.w.e.). The inner detector is a spherical acrylic vessel with a diameter of 35.4~m, containing 20~kton of linear alkylbenzene (LAB). The detector is instrumented with $\sim$17000 20-inch photomultiplier tubes and $\sim$25000 3-inch small photomultiplier tubes, enabling it to achieve an energy resolution of $\sim$3\% at 1 MeV~\cite{JUNO:2015zny}.

JUNO's primary goal is to determine the neutrino mass hierarchy through reactor neutrino oscillation measurements, which take advantage of JUNO's large target mass, low energy threshold, and superb energy resolution~\cite{JUNO:2022mxj}.  These also make JUNO well-suited for precision measurements of solar neutrinos through elastic neutrino-electron scattering~\cite{JUNO:2020hqc, JUNO:2023zty}.  JUNO's light yield is dominated by scintillation, which is isotropic, as opposed to Cherenkov emission, which is directional.  In Super-K, selecting events by their angle relative to the Sun reduces backgrounds by about an order of magnitude~\cite{Super-Kamiokande:2023jbt}. Because directional reconstruction at JUNO is still challenging --- plus the fact that it is shallower --- new techniques to reduce detector backgrounds are especially important.

At low energies (below 2.3~MeV), backgrounds from natural radioactivity are very large. These arise from detector materials, the surrounding rock, and the scintillator itself. The most relevant isotopes in this regard are $^{208}$Tl (originating from the glass) and the daughters of the decay chains of $^{238}$U, $^{232}$Th, and $^{222}$Rn (in the scintillator)~\cite{JUNO:2020hqc, JUNO:2023zty}. JUNO plans to mitigate these backgrounds using multiple purification techniques, energy-dependent fiducial volume (FV) cuts, and coincidence-based rejection of decay-chain isotopes~\cite{JUNO:2015zny, JUNO:2020hqc, JUNO:2023zty}.  With these cuts, JUNO is expected to achieve a world-leading energy threshold of $\sim$2~MeV in the innermost region of the detector. Compared to Borexino, JUNO's much larger size will allow for stronger fiducial volume cuts to eliminate $^{208}$Tl background decays, which were the limiting factor in setting Borexino's 3~MeV threshold~\cite{Borexino:2008fkj}.

Three spallation backgrounds are also important below 2.3 MeV. Two that are well known are $^{11}$C and $^7$Be, which have large yields and long lifetimes. The TFC cut mentioned above was designed specifically for $^{11}$C ~\cite{Galbiati:2004wx}. Below, we note that $^{13}$N is also important because of its long lifetime. We discuss these isotopes in more detail in Sec.~\ref{sec:Isotopes}, where we define our focus on spallation decays above 2.3~MeV (and with shorter lifetimes).

Above 2.3~MeV, other spallation backgrounds become the dominant challenge. Due to JUNO's relatively shallow depth, its muon rate is 5~s$^{-1}$~\cite{JUNO:2025fpc}, which is about two times larger than for Super-K (in the next paragraph, we explain this ratio).  To tag incoming muons, JUNO will employ a veto system consisting of a purified water volume surrounding the inner detector with photomultiplier tubes to detect the Cherenkov light, as well as an independent top tracker system to help with muon tagging and track reconstruction. Afterwards, cuts must be applied to eliminate any potential backgrounds caused by decays of isotopes produced through muon spallation. However, indiscriminate vetoes after all muons would impose excessive deadtime due to the high muon rate. Reducing this impact requires more refined cuts informed by the muon’s energy deposition and neutron yield, particularly those associated with hadronic showers, which produce almost all relevant spallation isotopes. As we explore in later sections, insights from our work about Super-K~\cite{Li:2014sea, Li:2015kpa, Nairat:2024upg} can guide the development of similar strategies in JUNO, adapted for its unique geometry and detection medium.

We now comment on the ratio of muon rates in JUNO and Super-K.  The muon flux at JUNO is $5.0 \times 10^{-7}$ cm$^{-2}$ s$^{-1}$~\cite{JUNO:2021vlw, JUNO:2025fpc}, while that at Super-K is $1.5 \times 10^{-7}$ cm$^{-2}$ s$^{-1}$~\cite{Super-Kamiokande:2015xra, Coffani:2021gbe}.  These values are closer than one might expect given the different depths of the two experiments, 1800 m.w.e.\ and 2700 m.w.e., respectively.  However, those are the vertical equivalent depths, as stressed in Ref.~\cite{Woodley:2024eln} (see their Fig.~4).  JUNO is under a relatively flat overburden, so we estimate the angle-averaged equivalent depth to be 1600 m.w.e.; in contrast, Super-K is under a mountain, with an angle-averaged equivalent depth of about 2100 m.w.e., which represents a larger change from the vertical case.  Additionally, taking into account that the inner-detector mass of JUNO is about 2/3 as large as that of Super-K, it makes sense that the muon rate in JUNO is 5~s$^{-1}$ while that in Super-K is 2~s$^{-1}$.

%%%%%%%%%%%%%%%%%%%%%%%%%%%%%%%%%%%%%%%%%%%%%%%%%%%%%%%%%%%%%%%%%%%%%%%

\subsection{Muon propagation in JUNO}

To study muon spallation at JUNO, we use the particle transport code \texttt{FLUKA} (version 4-5.0)~\cite{Ferrari:2005zk, Bohlen:2014buj, Battistoni:2015epi, Ahdida:2022gjl}, a Monte Carlo simulation tool that has been used extensively for spallation calculations and validated with experimental results~\cite{KamLAND:2009zwo,  Borexino:2013cke, Li:2014sea, Li:2015kpa, Li:2015lxa, Super-Kamiokande:2021snn, KamLAND-Zen:2023spw}.  We simulate the inner detector as a sphere of 35.4~m diameter. This is filled with LAB, which is not a default material in \texttt{FLUKA}, so we manually define it using the chemical composition C$_6$H$_5$C$_n$H$_{2n+1}$ (with $n=$ 10--13) and set its density to 0.86~g/cm$^3$. We take into account the natural abundance of different carbon isotopes. We then use the PRECISIOn defaults to activate all the relevant processes for muon propagation in matter, including ionization, delta-ray production, radiative energy losses, photonuclear interactions, and low-energy neutron transport.

\begin{figure}[t]
\includegraphics[width=0.99\columnwidth]{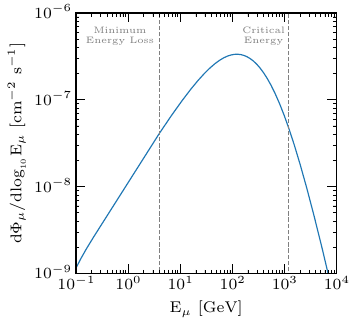}
\caption{Cosmic-ray muon flux at JUNO as a function of energy. The dashed lines represent the energy loss of a minimum ionizing muon and the critical energy of muons in linear alkylbenzene.}
\label{fig:MuonSpectrum}
\end{figure}

Figure~\ref{fig:MuonSpectrum} shows the muon energy spectrum at JUNO that we use as an input for our simulation. We estimate the spectrum in two ways, finding similar results. First, we extrapolate the preliminary JUNO simulations in Ref.~\cite{JUNO2014} to higher energies.  Second, we fit the general shape of the simulated muon energy distribution generated for other experiments and adjust the average energy to align with the expectations for JUNO.  For the average energy, we use 207~GeV~\cite{JUNO:2021vlw, Yufeng}. The exact spectrum requires a MUSIC~\cite{Antonioli:1997qw, Kudryavtsev:2008qh} simulation of cosmic-ray muons propagating underground to the detector, taking into account the geological map of the overburden. Currently, JUNO is conducting this detailed simulation, and the results are not yet publicly available~\cite{Yufeng}. However, our approach is reasonable because the detailed shape of the muon spectrum does not matter much (in fact, a delta function gives similar results).  The exception is the small fraction of isotopes made by stopping muons, for which we do need the spectrum shape.

%%%%%%%%%%%%%%%%%%%%%%%%%%%%%%%%%%%%%%
%%%%%%%%%%%%%%%%%%%%%%%%%%%%%%%%%%%%%%

The shape of the spectrum results from simple considerations.  We plot d$\Phi_\mu$/dlog$_{_{10}}$E$_{\mu}$ = $2.3\, $E$_\mu \,$d$\Phi_\mu$/dE$_\mu$ on the y-axis so that relative heights accurately represent the relative contributions to the total flux. With this choice, the spectrum rises as $\sim$$E_\mu$ at low energies because d$\Phi_\mu$/dE$_\mu$ is flat due to the large energy loss required to traverse the rock overburden. At high energies, E$_\mu\,$d$\Phi_\mu$/dE$_\mu$ falls as $\sim$$E_\mu^{-1.7}$ because the cosmic-ray muon spectrum falls as $\sim$$E_\mu^{-2.7}$. The normalization is set so that integration of the spectrum gives the expected muon flux at JUNO of $\sim$0.005~m$^{-2}$~s$^{-1}$, corresponding to a rate of $\sim$5~s$^{-1}$~\cite{JUNO:2021vlw, Yufeng, JUNO:2025fpc}.

We adopt some reasonable simplifications in our simulation. First, we simulate vertically down-going muons uniformly across the top of the detector,  which results in an average muon track length of 23.6~m, close to the 23~m value expected by JUNO simulations~\cite{JUNO:2015zny, JUNO:2020hqc}. Because our results are calculated per meter of muon track length, they can be easily rescaled for different track lengths of muons that are not exactly vertical.  (Further, the detector is spherically symmetric.)  Second, we only simulate $\mu^-$, but in reality, the muons entering the detector are a mixture of $\mu^-$ and $\mu^+$, with a ratio that can be measured in situ. However, the isotope yields of both $\mu^-$ and $\mu^+$ are almost identical except for those produced by nuclear captures of stopping $\mu^-$. Finally, we only simulate single muons because, with good muon reconstruction, muon bundles should be recognizable as messy events to be cut or can be treated as two individual muons.

With those simulation inputs, we use a modified user routine to calculate the energy loss, isotope and neutron yield, and kinetic energies of any daughter particles for each muon. Isotope yields are also checked with the built-in RESNUCLEi card. For the beta decay distributions, we use the RADDECAY card to activate the simulation of radioactive decays and a modified user routine that determines the time and energy of the decay products.

%%%%%%%%%%%%%%%%%%%%%%%%%%%%%%%%%%%%%%%%%%%%%%%%%%%%%%%%%%%%%%%%%%%%%%%

\subsection{Muon energy loss in JUNO}

As muons propagate through materials, they lose energy in two ways~\cite{Groom:2001kq, Dutta:2000hh, ParticleDataGroup:2024cfk}. The first is ionization and excitation of atomic electrons. Depending on the energy of the outgoing electrons, ionization losses can be separated into restricted ionization, which results in electrons with small outgoing energies, versus delta-ray production, which is characterized by hard collisions and thus higher-energy outgoing electrons. Restricted ionization happens at a continuous rate with small fluctuations, while delta-rays are produced through discrete interactions with large fluctuations.

The other way muons lose energy while propagating is through interactions with the atomic nuclei in the material via radiative processes. These include pair production, bremsstrahlung, and photonuclear interactions. Except for pair production, which leads to nearly continuous losses, radiative losses happen as discrete interactions with large fluctuations~\cite{Groom:2001kq, ParticleDataGroup:2024cfk}. This often leads to secondary particles with energies high enough to induce electromagnetic or hadronic showers. Almost all spallation isotopes are produced by these showers. 

In LAB, a minimum ionizing muon loses energy at a rate of $\sim$2~MeV\,g$^{-1}$\,cm$^2$. This corresponds to a minimum energy loss of $\sim$4~GeV by vertical through-going muons, which is represented by the left dashed line in Fig.~\ref{fig:MuonSpectrum}. Muons with kinetic energies below this threshold ($\sim$4\% of all muons) stop within the detector, undergoing either decay or nuclear capture. At higher energies, ionization losses increase slowly while radiative losses scale linearly with energy until both contributions become equal at the critical energy of 1.2~TeV, represented by the right dashed line in Fig.~\ref{fig:MuonSpectrum}. The fraction of muons with kinetic energy above this critical energy is $\sim$2\%.

At the muon average energy of 207~GeV, the energy losses are still dominated by ionization, but with a higher average loss rate of $\sim$2.8~MeV\,g$^{-1}$\,cm$^2$, plus $\sim$0.4~MeV\,g$^{-1}$\,cm$^2$ through radiative processes. While the radiative losses are smaller, they have large fluctuations that can lead to energetic secondary showers.

\begin{figure}[t]
    \includegraphics[width=0.99\columnwidth]{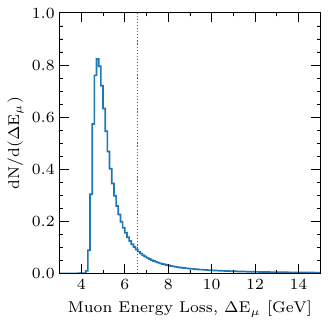}
\caption{Probability distribution of the total energy loss for through-going muons in JUNO, normalized per muon. The red dotted line corresponds to the average muon energy loss.}
\label{fig:MuonLoss}
\end{figure}

Figure~\ref{fig:MuonLoss} shows the total muon energy loss ($\Delta E_\mu$) probability distribution for the full spectrum of through-going muons at JUNO. We show the distribution for muons that travel the average track length of 23.6~m, but the result can be rescaled for any track length. The minimum energy loss of $\sim$4~GeV results from minimum ionizing muons. Stopping muons are not included in this distribution, though we take them into account in our other calculations. The most probable energy loss being at $\sim$5~GeV arises from the contributions of continuous processes (restricted ionization and pair production). The small fluctuations of these processes cause the narrow tail to the left. The large fluctuations in delta-ray production, bremsstrahlung, and photonuclear interactions cause the broad tail to the right. On average, muons lose a total energy of $\sim$6.5~GeV, corresponding to the average energy loss rate of $\sim$3.2~MeV\,g$^{-1}$\,cm$^2$.

The muon-induced shower energy is defined as the muon energy loss minus the energy that would be lost by a minimum ionizing muon, though this has some subtleties~\cite{Nairat:2024upg}. This is equivalent to the total energy deposited into secondary particles (electrons, positrons, gamma rays, neutrons, or charged pions) that can induce electromagnetic or hadronic showers. Muons with higher shower energy produce more secondary particles and are, therefore, more likely to produce spallation isotopes.

%%%%%%%%%%%%%%%%%%%%%%%%%%%%%%%%%%%%%%
%%%%%%%%%%%%%%%%%%%%%%%%%%%%%%%%%%%%%%

%%%%%%%%%%%%%%%%%%%%%%%%%%%%%%%%%%%%%%%%%%%%%%%%%%%%%%%%%%%%%%%%%%%%%%%
%%%%%%%%%%%%%%%%%%%%%%%%%%%%%%%%%%%%%%%%%%%%%%%%%%%%%%%%%%%%%%%%%%%%%%%

\section{Spallation Backgrounds in JUNO}
\label{sec:Isotopes}

In this section, we calculate the predicted yields of isotopes produced by muon spallation in JUNO, their decay energy spectra, and their time distributions. These are a prerequisite to our main new results for improving spallation cuts.

%%%%%%%%%%%%%%%%%%%%%%%%%%%%%%%%%%%%%%%%%%%%%%%%%%%%%%%%%%%%%%%%%%%%%%%

\subsection{Predicted isotope yields at JUNO}

%%%%%%%%%%%%%%%%%%%%%%%%%%%%%%%%%%%%%%
\begin{table*}[!t]
\renewcommand{\arraystretch}{1.3}

\begin{tabular}{cccccc}
\toprule
Isotope & Q value [MeV] & Half-life [s] & Decay mode & Yield [$10^{-7}$$\mu^{-1}$g$^{-1}$cm$^2]$ & Primary process\\
\midrule

$^{12}$N    &  17.3  &  0.011  &  $\beta^+$  &   0.51   & (p,n)  \\

$^{10}$C    &  3.65  &  19.3  &  $\beta^+\gamma$  &   8.43   & ($\pi^+$,n+p)  \\

$^{9}$C    &  16.5  &  0.127  &   $\beta^+$ &   0.72  &  ($\pi^+$,2n+p) \\

$^{13}$B    &  13.4  &   0.0174   &  $\beta^-$ &   0.22   &  $^{13}$C(n,p) \\

$^{12}$B    &  13.4  &  0.0202  &  $\beta^-$  &   22.24   & (n,p)  \\

$^{8}$B    &  18.0  &  0.770  &  $\beta^+ \alpha$  &   5.13   &  ($\pi^+$,2n+2p)  \\

$^{11}$Be    &  11.5  &  13.76  &  $\beta^-$   &   1.43   &  (n,2p)  \\

$^{9}$Li    & 13.6   &  0.178   &  $\beta^-$ (49.5\%) $\beta^- n$ (50.5\%)  &   2.06   & (n,2p+$^2$H)  \\

$^{8}$Li    &  16.0  &  0.840   & $\beta^-\alpha$   &   17.36   &  (n,p+$\alpha$) \\

$^{8}$He    &  10.7  &  0.119  &  $\beta^-$ (84\%)  $\beta^- n$ (16\%)  &  0.20   & ($\pi^-$,2p+$^2$H)   \\

$^{6}$He    &  3.51  & 0.807   &  $\beta^-$  &  10.92   &  ($\pi^-$,n+p+$\alpha$) \\
\midrule
$^{13}$N    &  2.22  & 598   & $\beta^+$   &   0.93   &  ($^2$H,n) \\
$^{11}$C & 1.98 & 1222 & $\beta^+$ & 440 & ($\gamma$,n) \\
$^{7}$Be & 0.86 & $4.6 \times 10^6$ & EC & 94.31 & ($\gamma$,n+$\alpha$) \\

\midrule
$^{11}$B    &    &   &   &   740   &  ($\gamma$,p) \\

$^{10}$B    &    &   &   &   120   &  (n,2n+p) \\

$^{10}$Be    &    &   &   &   27   &  (n,n+2p) \\

$^{9}$Be    &    &   &   &   107   &  ($\gamma$,$^3$He) \\

$^{7}$Li    &    &   &   &   153   &  (n,n+p+$\alpha$) \\

$^{6}$Li    &    &   &   &   168   &  (n,2n+p+$\alpha$) \\

\midrule
sum & & & & 1919 &\\
n & & & & 1665\\
\bottomrule
\end{tabular}
\caption{Predicted isotope yields in JUNO. The top block includes unstable isotopes with decay energies above 2.3~MeV.  The second block includes isotopes with decay energy endpoints below 2.3~MeV. The third block contains stable isotopes that do not cause backgrounds. The last block shows the sum of all isotope yields and the yield of spallation neutrons. Decay energies, half-lives, and decay modes of unstable isotopes are taken from Ref.~\cite{NNDC}. The yields and primary production processes are from our \texttt{FLUKA} simulation. Yields are expressed in units of $\mu^{-1}$\,g$^{-1}$\,cm$^2$. The conversion factor from this unit to day$^{-1}$~kton$^{-1}$ is $\sim$4.3$\times 10^7$, assuming a muon rate of $5.0 \times 10^{-7}$ cm$^{-2}$.}
\label{table:Yields}
\end{table*}
%%%%%%%%%%%%%%%%%%%%%%%%%%%%%%%%%%%%%%

We start by calculating the yields of all unstable isotopes produced as muons propagate through the detector. These isotopes are almost all produced by interactions of secondary particles generated by the various muon energy loss processes, which we discuss in detail in the next section. We calculate the expected yields using \texttt{FLUKA} by propagating muons and simulating their energy loss processes. We then propagate all the secondary particles produced through these energy losses and simulate their interaction processes that produce isotopes. The yield of an isotope produced in this manner is proportional to the convolution of the path length spectra of the secondary particles and the cross section for the various isotope production processes~\cite{Galbiati:2004wx, Galbiati:2005ft}.

Table~\ref{table:Yields} shows the yields for all unstable (upper two blocks) and stable isotopes (third block) produced per muon, along with the decay mode, energy, and half-life of the unstable isotopes. It also shows the primary production process of each isotope. The yields are reported in units of $\mu^{-1}$\,g$^{-1}$\,cm$^2$, obtained by dividing the yield per muon by the target material's density and the muon track length. 

The sum of all isotope yields in JUNO, which is 1919$\times 10^{-7}$~$\mu^{-1}$\,g$^{-1}$\,cm$^2$, corresponds to an average of 0.4 spallation isotopes per muon. The neutron yield is comparable, $\sim$0.34 per muon. Only 31\% of the spallation isotopes are unstable with a total yield of 604$\times 10^{-7}$~$\mu^{-1}$\,g$^{-1}$\,cm$^2$, corresponding to 0.12 per muon. With the expected muon rate of 5~s$^{-1}$, this means a rate of $\sim$0.6 unstable isotopes produced in the detector per second. 

The unstable isotopes with the largest yields are by far $^{11}$C and $^7$Be, which are primarily produced through gamma-ray interactions. Despite the lower cross sections of these interactions relative to other processes~\cite{Brown:2018jhj, Koning:2019qbo}, the high abundance of gamma rays in electromagnetic showers frequently induced by muons results in large yields of these isotopes. 

Among the spallation products, $^{11}$C and $^7$Be also stand out for their large yields and the longest lifetimes.  Considering the muon rate of 5~s$^{-1}$, imposing a cut after each muon for a duration of $\sim$10$^3$~s (the shorter half-life of the two) would effectively blind the detector. Consequently, the decay energy endpoints of these two isotopes directly affect JUNO's solar neutrino sensitivity. The $^{11}$C beta-decay spectrum ends at 1.98~MeV, while 10.7\% of $^7$Be electron-capture decays result in gamma rays with an energy of 0.477~MeV~\cite{NNDC}. 

Another isotope of concern is $^{13}$N. Despite its relatively smaller yield, with a half-life of 598~s, it would also be difficult to reduce this background without severe exposure loss. Its beta decay spectrum ends at 2.22 MeV, or $\sim$2.3~MeV, with the energy resolution taken into account. This isotope was not considered in Ref.~\cite{JUNO:2020hqc}, and we found no direct yield measurements of it in previous experiments. KamLAND, however, reported a simulated yield of 0.06$\times 10^{-7}$~$\mu^{-1}$\,g$^{-1}$\,cm$^2$ using \texttt{FLUKA}, with the primary process of production being $^{13}$C(p,n)$^{13}$N~\cite{KamLAND:2009zwo}. Our simulation, by contrast, gives a much larger yield of 0.93$\times 10^{-7}$~$\mu^{-1}$\,g$^{-1}$\,cm$^2$, dominated by the process $^{12}$C(d,n)$^{13}$N. 

We attribute this discrepancy to the treatment of light ions, specifically deuterons, in the \texttt{FLUKA} simulation. When we disable light-ion transport in our simulation, we recover the results of Ref.~\cite{KamLAND:2009zwo}. This may be related to updates in the implementation of light-ion transport in newer versions of \texttt{FLUKA}. The only other isotope affected by this is $^{13}$C, which can be produced through (d,p) reactions. However, $^{13}$C is stable and does not contribute to the backgrounds.

Taken together, $^{11}$C, $^{7}$Be, $^{13}$N, and contributions from natural radioactivity~\cite{JUNO:2020hqc} overwhelm the sub-2.3 MeV region. While Ref.~\cite{JUNO:2020hqc} suggested setting the $^8$B solar neutrino analysis threshold at 2.0 MeV, we raise it to 2.3 MeV in this paper to account for $^{13}$N. Even so, such a measurement by JUNO would represent the lowest threshold ever reached for $^8$B solar neutrino detection.

Apart from $^{11}$C and $^7$Be, the primary production process of the unstable isotopes involves neutrons, pions, or light ions from the rare hadronic showers, which explains their relatively lower yields. The total yield of isotopes with decay energies larger than 2.3~MeV is 69$\times 10^{-7}$~$\mu^{-1}$\,g$^{-1}$\,cm$^2$, however, a fraction of their decays yield a visible energy below the 2.3~MeV threshold. For example, only 18\% of $^6$He decays are above the threshold. Overall, the rate of background events above the threshold is 55$\times 10^{-7}$~$\mu^{-1}$\,g$^{-1}$\,cm$^2$, corresponding to 0.011 isotope per muon, or 4800 isotope decays per day. For comparison, the expected rate of $^8$B solar neutrino signals in this energy range at JUNO is less than 100 events per day, which shows the significance of spallation backgrounds. 

Negative muons with low enough energies to stop inside the detector lead to the production of unstable isotopes after the muon capture on carbon nuclei~\cite{Galbiati:2005ft}. These muons are responsible for 5\% of all isotopes with decay energies above 2.3~MeV. The two major contributions of stopping muons are the production of $^{12}$B and $^{8}$Li through ($\mu^-$,$\nu_\mu$) and ($\mu^-$,$\alpha$+$\nu_\mu$). The fractions of $^{12}$B and $^{8}$Li isotopes produced by stopping muons are about 10\% and 5\% of their yields, respectively. 

Our predicted yields are generally consistent with those for other scintillator-based neutrino experiments like KamLAND and Borexino~\cite{KamLAND:2009zwo, Borexino:2013cke}. Our yields are also consistent with the recent measurements in KamLAND-Zen~\cite{KamLAND-Zen:2023spw} because the xenon loading is only a few percent by mass and has a negligible effect on the rate of relevant spallation products. JUNO has a shallower depth compared to these experiments, resulting in a lower average muon energy, so our yields are expected to be slightly smaller. Overall, most yields agree within a factor of two, which is reasonable considering hadronic uncertainties and the difference in muon average energies. These uncertainties do not directly impact the main results of this paper, which are the background rejection efficiency and deadtime reduction achieved by the proposed cuts.

We can also compare the yields to those in Super-K because all of these isotopes are also produced by muon spallation in water Cherenkov detectors. The sum of all isotope yields in Super-K is $\sim$3000$\times 10^{-7}$~$\mu^{-1}$\,g$^{-1}$\,cm$^2$~\cite{Li:2014sea}, compared to $\sim$1900$\times 10^{-7}$~$\mu^{-1}$\,g$^{-1}$\,cm$^2$ in JUNO. The larger yield at Super-K is expected due to the higher average muon energy compared to JUNO (271~GeV compared to 207~GeV). Above their respective detection thresholds, the sum of background isotope yields is $\sim$50$\times 10^{-7}$~$\mu^{-1}$\,g$^{-1}$\,cm$^2$ in Super-K~\cite{Li:2014sea}, compared to $\sim$70$\times 10^{-7}$~$\mu^{-1}$\,g$^{-1}$\,cm$^2$ in JUNO, which is also expected due to the lower energy threshold at JUNO. The yields of individual isotopes are also comparable when considered in terms of the production process. For instance, the yield of $^{12}$B in scintillator detectors (22$\times 10^{-7}$~$\mu^{-1}$\,g$^{-1}$\,cm$^2$) and $^{16}$N in water Cherenkov detectors (18$\times 10^{-7}$~$\mu^{-1}$\,g$^{-1}$\,cm$^2$) are close because they are both primarily produced through (n,p) interactions.

%%%%%%%%%%%%%%%%%%%%%%%%%%%%%%%%%%%%%%%%%%%%%%%%%%%%%%%%%%%%%%%%%%%%%%%

\subsection{Spallation spectra and time distributions}

%%%%%%%%%%%%%%%%%%%%%%%%%%%%%%%%%%%%%%
\begin{figure*}[t]
    \includegraphics[width=0.98\textwidth]{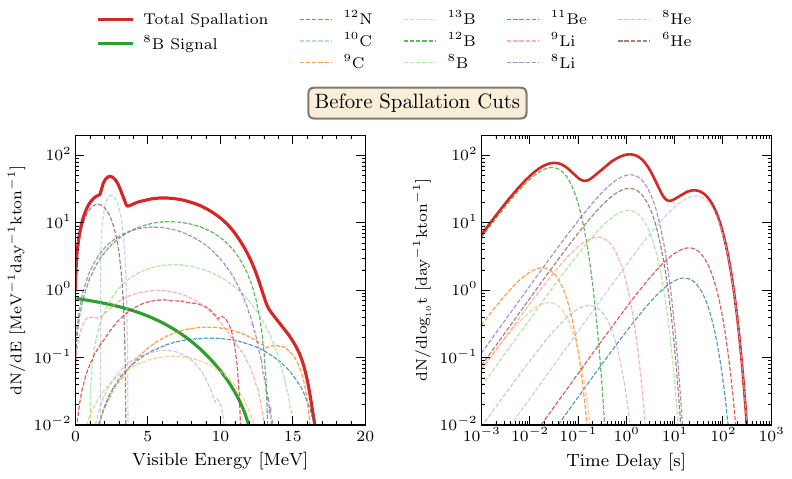}
    \includegraphics[width=0.98\textwidth]{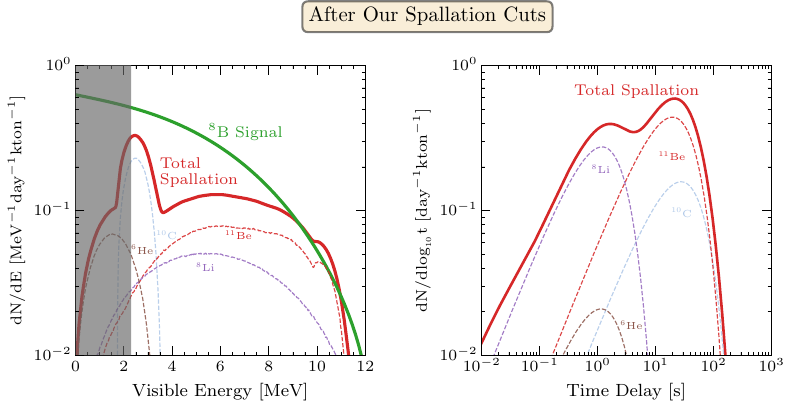}
\caption{Spallation background energy spectra and time distributions, before and after our cuts (note the changes in the axes ranges). The expected energy spectrum of the $^8$B solar neutrino signal (taken from Ref.~\cite{JUNO:2020hqc}) is shown for comparison.}
\label{fig:SpallationSpectra}
\end{figure*}

%%%%%%%%%%%%%%%%%%%%%%%%%%%%%%%%%%%%%%

We now calculate the energy spectra and time distributions for the unstable spallation isotopes. Using \texttt{FLUKA} to simulate radioactive decays, we calculate the visible energy for each decay by summing the kinetic energies of the produced betas and gammas (which deposit most of their energy through Compton scattering and pair production). For positrons, we include the annihilation energy. The energy of alpha particles is not included because their visible energies after quenching are usually below the detection threshold~\cite{JUNO:2020hqc}. The energy released by neutron captures is also not included because they occur after a relatively long delay. The decay spectra obtained by the simulation were cross-checked and found to agree with the nuclear data from Ref.~\cite{NNDC}, except $^6$He, which is not simulated in \texttt{FLUKA}; for that, we use the spectrum from Ref.~\cite{NNDC}.

Figure~\ref{fig:SpallationSpectra} (top left panel) shows the total energy spectrum for the decays of unstable spallation isotopes with a decay energy endpoint above the 2.3~MeV threshold (smeared assuming an energy resolution of 3\%/$\sqrt{E\,(\text{MeV})}$~\cite{JUNO:2024fdc}). We also show the individual spectra of the isotopes that contribute significantly to the total spectrum. Each of these is weighted by the isotope's predicted yield, so their normalization corresponds to their yields in Table~\ref{table:Yields} with the units converted to day$^{-1}$~kton$^{-1}$. For comparison, we also show the expected spectrum of $^8$B neutrino-electron elastic scattering signals at JUNO, taken from Ref.~\cite{JUNO:2020hqc}. Before cuts, the signal is 10--100 times smaller than the spallation background. 

Figure~\ref{fig:SpallationSpectra} (top right panel) shows the decay time distribution of the spallation isotopes weighted by their yields. We plot dN/dlog$_{_{10}}$t such that the area under the curve correctly represents the integral, as explained for Fig.~\ref{fig:MuonSpectrum}. The time delay here is calculated relative to the individual muon that produced the isotope. To reject $\sim$90\% of the decay backgrounds caused by a certain isotope, the cuts after the muons need a rejection period of $\sim$2~times the lifetime of that isotope. Therefore, the longer the lifetime of the isotope, the more deadtime the usual muon vetoes for that isotope would result in.

Here and below, we make predictions for a representative 1-kton mini-volume of JUNO.  For full-detector results, one would have to account for all of the mini-volumes included in the energy-dependent fiducial volume~\cite{JUNO:2020hqc}.  Above 5~MeV, JUNO expects a fiducial volume of 16.2~kton; between 3 and 5~MeV, they expect 12.2~kton; and between 2 and 3~MeV, they expect 7.9~kton.  The inner-detector volume is 20~kton.

%%%%%%%%%%%%%%%%%%%%%%%%%%%%%%%%%%%%%%%%%%%%%%%%%%%%%%%%%%%%%%%%%%%%%%%
%%%%%%%%%%%%%%%%%%%%%%%%%%%%%%%%%%%%%%%%%%%%%%%%%%%%%%%%%%%%%%%%%%%%%%%

\section{Shower Production in JUNO}
\label{sec:Showers}

In this section, we show how muon energy loss results in the production of secondary showers, neutrons, and isotopes.  For further details with similar underlying physics, see our paper about Super-K~\cite{Nairat:2024upg}.

%%%%%%%%%%%%%%%%%%%%%%%%%%%%%%%%%%%%%%%%%%%%%%%%%%%%%%%%%%%%%%%%%%%%%%%

\subsection{Secondary particle showers}

%%%%%%%%%%%%%%%%%%%%%%%%%%%%%%%%%%%%%%
\begin{figure}[t]
    \includegraphics[width=0.99\columnwidth]{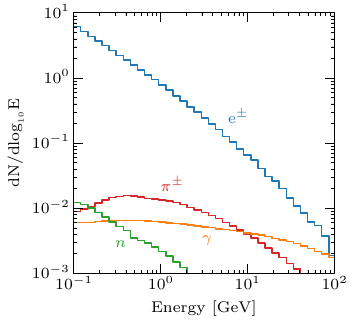}
\caption{The energy spectrum of secondary particles produced directly by the muons in JUNO, normalized per muon.}
\label{fig:Secondaries}
\end{figure}
%%%%%%%%%%%%%%%%%%%%%%%%%%%%%%%%%%%%%%
%%%%%%%%%%%%%%%%%%%%%%%%%%%%%%%%%%%%%%
\begin{figure*}[t]
    \includegraphics[width=0.96\textwidth]{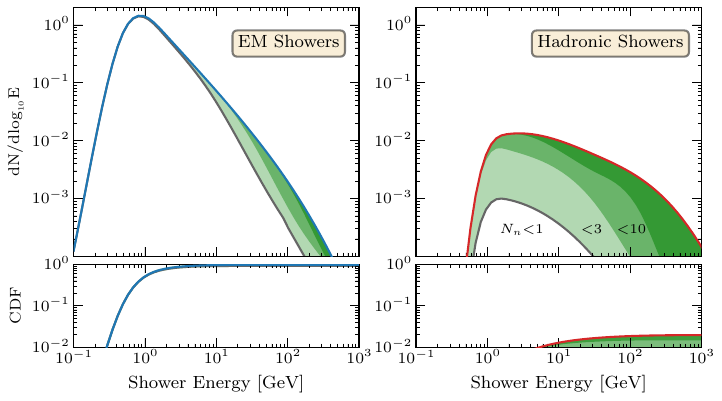}
    \includegraphics[width=0.96\textwidth]{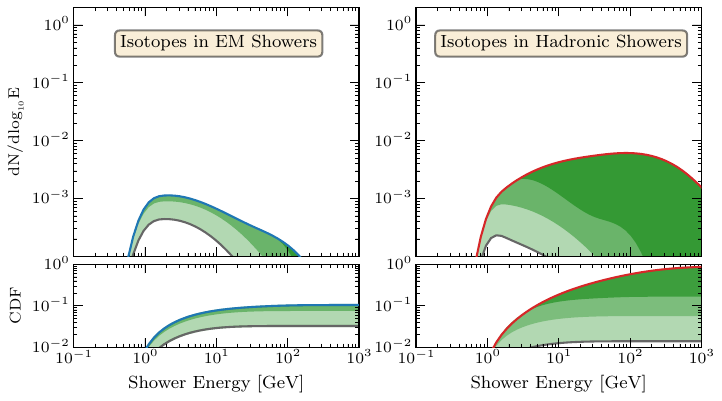}
\caption{Energy spectra of electromagnetic and hadronic showers induced by muons in JUNO. The top panels show the energy spectra for all showers, normalized per muon. The bottom panel shows the spectra of isotope-producing showers only, weighted by their isotope yields. In all panels, the different shades represent the associated neutron yields as shown in the top right panel.}
\label{fig:Showers}
\end{figure*}
%%%%%%%%%%%%%%%%%%%%%%%%%%%%%%%%%%%%%%

To show how muon energy losses lead to secondary showers, we start by calculating the energy spectrum of daughter particles produced directly by muons. These particles are primarily electrons from delta-ray production, positrons from pair production, gamma rays from bremsstrahlung interactions, and pions or neutrons from photonuclear interactions.

Figure~\ref{fig:Secondaries} shows the kinetic energy spectrum of the daughter particles produced by muons, normalized per muon. The frequency of electrons and positrons is much larger than that of other particles because muon energy losses are dominated by ionization, delta-ray production, and pair production. Gamma rays are produced through bremsstrahlung interactions. Pions and neutrons are produced through photonuclear interactions.

The shape of these spectra follows from the differential cross section of the production processes, as detailed in Refs.~\cite{Li:2015kpa, ParticleDataGroup:2024cfk}. Compared to the results for Super-K in Ref.~\cite{Li:2015kpa}, there are slight differences that arise because of the difference in average muon energy. Ionization does not change much with muon energy, so the overall rate of daughter electrons and positrons is comparable. Pions and gamma rays have a lower rate here than Super-K because radiative losses scale linearly with energy. 

The rate of neutrons produced directly by muons is $\sim$0.05 neutrons per through-going muon, which is only 15\% of the total neutron production rate in Table~\ref{table:Yields}. The majority of neutrons are produced in secondary showers rather than the muons themselves.

Secondary showers are formed when the energy of the daughter particles exceeds their critical energy in LAB. For electrons and positrons, this energy is about 100~MeV~\cite{Groom:2001kq,ParticleDataGroup:2024cfk}. Any secondary electrons or positrons with higher initial energy are likely to initiate an electromagnetic shower, characterized by multiple generations of electrons, positrons, and gamma rays, through bremsstrahlung and pair production processes. Similarly, charged pions can initiate hadronic showers through a succession of inelastic collisions with nuclei. However, the critical energy of hadronic showers is about 10 times higher than that of electromagnetic showers, which makes them less frequent.

In principle, each daughter particle in Fig.~\ref{fig:Secondaries} with enough energy can initiate an individual shower, electromagnetic if the particle is an electron, positron, or gamma ray, and hadronic if it is a charged pion. The shower energy is the kinetic energy of that initial particle. However, in practice, individual showers cannot always be identified. As noted above, we define the shower energy as the total muon energy loss above minimum ionization, which can often be the sum of multiple individual showers. The results in this paper are not dependent on the exact energy scale, so in the following, we define the shower energy as the total muon energy loss above its minimum ionization loss rather than the sum energy of individual showers.

%%%%%%%%%%%%%%%%%%%%%%%%%%%%%%%%%%%%%%%%%%%%%%%%%%%%%%%%%%%%%%%%%%%%%%%

\subsection{Electromagnetic versus hadronic showers: neutron and isotope production}

As showers develop and the number of secondary particles grows larger, the probability of isotope production increases. Separating electromagnetic and hadronic showers is crucial because the secondary particles in these showers differ, leading to different isotope production mechanisms. Additionally, there is a difference in the geometry of electromagnetic and hadronic showers, necessitating different cuts for each type of shower.

In electromagnetic showers, isotopes are primarily produced through gamma-ray interactions. Due to the high frequency of electromagnetic showers, the backgrounds caused by these isotopes are difficult to reduce without causing a large deadtime. As mentioned previously, only two of the unstable isotopes in JUNO are primarily produced in this manner, $^{11}$C and $^{7}$Be, both of which have a low decay energy and can be eliminated using an energy threshold cut. 

In hadronic showers, the major parent particles for isotope production are neutrons and charged pions. One of the key characteristics of hadronic showers is the abundance of neutrons. Hadronic showers generally produce $\sim$10 times more neutrons than electromagnetic showers of the same energy. A small fraction of these neutrons interact and produce isotopes, but most of them scatter down to thermal energies and capture on nuclei. In LAB, the vast majority of thermal neutrons capture on hydrogen, emitting 2.2~MeV gamma rays. These captures are detectable in JUNO with high efficiency, providing a powerful tool to tag hadronic showers.

Figure~\ref{fig:Showers} (top panel) shows the energy distribution of electromagnetic and hadronic showers and the relative contribution of showers with and without neutrons, with the different shades of green representing the neutron multiplicity. The distributions are normalized per muon, so the cumulative distribution functions underneath represent the ratio of muons with showers under a certain energy to all muons. Because the shower energy is defined as the energy loss above minimum ionization, these distributions essentially do not include stopping muons. A shower is classified as hadronic if at least one charged pion or kaon is produced in it. While this definition is not exact because electromagnetic showers can occasionally develop a hadronic component, it is enough to separate the showers into two categories. 

Electromagnetic showers constitute 98\% of all muon-induced showers, but only 6\% of them yield neutrons. Conversely, only 2\% of muons generate hadronic showers, and 95\% of those result in at least one detectable neutron capture. The average number of neutrons produced by muons with electromagnetic showers is 0.08, compared to 13.5 for muons with hadronic showers. Overall, $\sim$8\% of all muons produce neutrons, which means that neutron-based cuts would cause minimal deadtime.

Figure~\ref{fig:Showers} (bottom panel) shows similar energy distributions of the showers weighted by their isotope yields. We only include isotopes with decay energies above 2.3~MeV. The cumulative distribution functions show the relative fraction of isotopes produced by showers under a certain energy to the total number of isotopes.

While electromagnetic showers dominate the muon energy losses and account for the majority of muon-induced showers, they only produce 10\% of the background isotopes. Rare hadronic showers dominate isotope production and produce the remaining 90\%. Because almost all hadronic showers (and some electromagnetic showers) produce neutrons, 96\% of background isotopes are accompanied by at least one neutron capture, and over 85\% of them are accompanied by more than three neutron captures. This demonstrates the high efficiency of neutron-based cuts in reducing backgrounds despite their low deadtime cost.

It is important to note that the neutron yields discussed here originate in two ways. First, neutrons can be produced directly in the isotope production processes, as for reactions like $^{12}$C($\pi^+$,n+p)$^{10}$C and $^{12}$C($\gamma$,n)$^{11}$C.  These are taken into account, though they are only a small fraction of the total yield of neutrons.  Second, the majority of neutrons are produced during the development of hadronic showers, preceding isotope production. Despite their smaller neutron yields, electromagnetic showers still produce a considerable fraction  ($\sim$23\%) of the total neutron yield due to their high frequency~\cite{Nairat:2024upg}.  As a point of clarification, we do not consider neutrons that can occur in some beta decays, e.g., from $^9$Li and $^8$He, because those come after the spallation decay, not before.

TFC cuts are optimized for the first type: neutrons produced alongside isotopes. The small spherical veto regions allow long time windows, making them particularly effective at rejecting long-lived isotopes such as $^{11}$C and $^{10}$C~\cite{Galbiati:2004wx, Borexino:2011cjz, Borexino:2021pyz, KamLAND-Zen:2023spw}. However, TFC cuts alone cannot remove all shower-induced isotopes, because many are produced without an accompanying neutron --- or can consume neutrons through processes like (n,p) reactions --- and can therefore appear far from any neutron capture point. As a representative case, consider a representative hadronic shower around a muon track: a $^{10}$C isotope produced 2~m to the right via ($\pi^+$,n+p) would be removed by a TFC cut, while a neutron 2~m to the left undergoes (n,p) to create $^{12}$B, which TFC would miss.

Cylindrical cuts address this limitation of TFC cuts by tagging hadronic showers and vetoing the full shower region, thereby ensuring containment of all associated isotopes. TFC vetoes remain crucial, however, for efficiently rejecting long-lived isotopes such as $^{11}$C and $^{10}$C. Together, they provide complementary coverage, achieving high background rejection with minimal deadtime.

%%%%%%%%%%%%%%%%%%%%%%%%%%%%%%%%%%%%%%%%%%%%%%%%%%%%%%%%%%%%%%%%%%%%%%%

\subsection{Spatial distribution of isotopes}

%%%%%%%%%%%%%%%%%%%%%%%%%%%%%%%%%%%%%%

\begin{figure*}[t]
    \includegraphics[width=0.99\columnwidth]{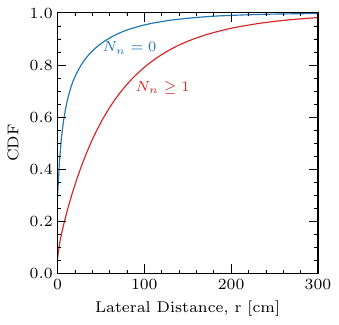}
    \includegraphics[width=0.99\columnwidth]{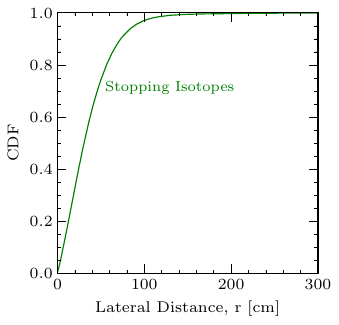}
\caption{Cumulative distributions of background isotopes as a function of their lateral displacement from the muon track. The left panel shows the isotopes produced by through-going muons, with and without neutrons. The right panel is for isotopes produced by stopping muons. In both panels, we do not count isotopes with decay energies below the threshold, such as $^{11}$C.}
\label{fig:Distances}
\end{figure*}

%%%%%%%%%%%%%%%%%%%%%%%%%%%%%%%%%%%%%%

Because isotopes are produced in showers, the necessary size of the cuts depends on the geometry of these showers. The longitudinal and lateral profiles of showers have been studied extensively in Refs.~\cite{Li:2015kpa, Nairat:2024upg}. Generally, the longitudinal extent of showers grows with energy and is roughly the same for both electromagnetic and hadronic showers, because the hadronic interaction length is comparable to the radiation length in this low-$Z$ material. However, the lateral extent of electromagnetic showers is smaller than that of hadronic showers.

Figure~\ref{fig:Distances} (left panel) shows the cumulative distribution function for the lateral displacement of background produced by through-going muons relative to their track. To show the difference between electromagnetic showers and hadronic showers, we separate the contribution of showers without neutrons (which are mostly electromagnetic) and with neutrons (hadronic). An important exception to this behavior is $^{11}$C, which despite being produced in electromagnetic showers, is usually accompanied by neutrons. However, $^{11}$C is not included in Fig.~\ref{fig:Distances} because its decay energy is below the threshold. The right panel shows a similar plot for the displacement of isotopes produced by stopping muon captures relative to the muon entry point. This displacement is caused by the multiple scattering of low-energy muons as they come to rest.

Almost all isotopes produced in showers with neutrons are contained within a $\sim$3.5~m radius cylinder around the muon track. On the other hand, isotopes produced without neutrons are almost all contained within a $\sim$1.5~m radius around the muon track because they tend to originate in electromagnetic showers. For stopping isotopes, a cylindrical cut with a radius of $\sim$1.5~m would be enough to contain almost all of them.

The key findings of our results are thus as follows. The majority of background isotopes are produced by hadronic showers. Neutrons are produced abundantly in hadronic showers and rarely in electromagnetic showers, making them a reliable indicator of the shower type. Last, hadronic showers occur at a low frequency, allowing for large-sized cuts (3--6~m radius) to be applied to them. In contrast, the remaining majority of showers are electromagnetic and only require cuts of smaller extent (1--3~m radius).

%%%%%%%%%%%%%%%%%%%%%%%%%%%%%%%%%%%%%%%%%%%%%%%%%%%%%%%%%%%%%%%%%%%%%%%
%%%%%%%%%%%%%%%%%%%%%%%%%%%%%%%%%%%%%%%%%%%%%%%%%%%%%%%%%%%%%%%%%%%%%%%

\section{Spallation Cuts}
\label{sec:Cuts}

In this section, we propose some improvements to spallation cuts at JUNO motivated by our presented results.

%%%%%%%%%%%%%%%%%%%%%%%%%%%%%%%%%%%%%%%%%%%%%%%%%%%%%%%%%%%%%%%%%%%%%%%

\subsection{JUNO’s planned cuts}

JUNO’s planned background reduction strategy is based on a sequence of cuts designed to target different backgrounds. Their first step is setting the analysis energy threshold to 2~MeV, which removes backgrounds from isotopes such as $^{11}$C and $^{7}$Be, as well as the most important natural radioactivity backgrounds. Energy-dependent fiducial volume cuts are also applied to further suppress backgrounds from detector materials to a subdominant level above 2~MeV \cite{JUNO:2020hqc}. As noted above, we think this threshold should be increased to 2.3~MeV.

Reducing the remaining spallation backgrounds relies on muon track reconstruction. Cosmic-ray muons entering the detector can be classified into single through-going muons, muon bundles, and stopping muons. Single through-going muons can be reconstructed with high efficiency~\cite{JUNO:2025gmd}, and muon bundles can often be resolved into one or two tracks depending on their separation~\cite{Genster:2018caz, Zhang:2018kag, Wonsak:2018uby, Liu:2021okf, Yang:2022din, JUNO:2025gmd}. However, in cases where track reconstruction fails --- such as for extremely high-energy muons ($>100$~GeV) that saturate the detector --- JUNO applies a full-volume veto, which contributes negligibly to deadtime given the rarity of such events~\cite{JUNO:2020hqc}.

For reconstructed muons, JUNO plans to apply concentric cylindrical cuts with radii up to 5~m and time windows up to 5~s~\cite{JUNO:2020hqc}, removing all events in the following ranges:
\begin{itemize}
    \item $r<$~1~m for 5~s,
    \item 1~m~$< r <$~3~m for 4~s,
    \item 3~m~$< r <$~4~m for 2~s,
    \item 4~m~$< r <$~5~m for 0.2~s.
\end{itemize}
These cuts are applied uniformly to all muons, regardless of whether or not they produce neutrons. While effective at removing spallation isotopes, this approach introduces substantial deadtime of 56\%(originally estimated as 44\% in Ref.~\cite{JUNO:2020hqc} because the muon rate at JUNO was initially expected to be 3.6~s$^{-1}$ instead of 5~s$^{-1}$). Furthermore, a spherical TFC cut with a radius of 2~m is applied around neutron captures for 160~s, causing an additional 6\% deadtime.

Stopping muons are a challenge. Unlike through-going muons, stopping muons lack an exit point, making their reconstruction more difficult. Nonetheless, various reconstruction algorithms demonstrated the ability to reconstruct straight-line tracks or localize the stopping point using scintillation light with sufficient accuracy~\cite{Wonsak:2018uby, Yang:2023rbg}. While Ref.~\cite{JUNO:2020hqc} does not specifically address stopping muons, we assume that they can be well-reconstructed, and the above-mentioned cylindrical cuts are applied to them.

%%%%%%%%%%%%%%%%%%%%%%%%%%%%%%%%%%%%%%%%%%%%%%%%%%%%%%%%%%%%%%%%%%%%%%%

\subsection{Proposed improved cuts}

We propose a refinement of JUNO’s spallation cut strategy by separating muons into distinct categories --- neutron-producing through-going muons (hadronic showers), non-neutron through-going muons (electromagnetic showers), and stopping muons --- and applying optimized cuts to each class. This approach exploits the physical differences between shower types, allowing aggressive cuts where justified while avoiding excessive deadtime. We also set the threshold to 2.3~MeV.

We separate through-going muons into two categories based on their neutron production. Neutron-producing muons represent 7.8\% of through-going muons, but account for 95.3\% of their background spallation products. For these, we apply cylindrical cuts as follows:
\begin{itemize}
    \item $r<$~0.5~m for 40~s,
    \item 0.5~m~$< r <$~4~m for 4~s,
    \item 4~m~$< r <$~6~m for 0.2~s.
\end{itemize}
These cuts cause only $\sim$7\% deadtime. By applying these stronger cuts to a smaller subset of muons, the deadtime cost is significantly reduced compared to JUNO's baseline approach.  In addition to these cylindrical cuts, we also apply the same spherical TFC cuts proposed in Ref.~\cite{JUNO:2020hqc} to reduce $^{10}$C, which adds an additional $\sim$5\% deadtime. Together, these cuts reduce the fraction of backgrounds that are produced with neutrons to $\sim$0.2\% of the original total.

For through-going muons without neutrons, which produce the remaining 4.7\% of isotopes, we apply smaller cylindrical cuts as follows:
\begin{itemize}
    \item $r<$~1~m for 5~s,
    \item 1~m~$< r <$~3~m for 0.2~s.
\end{itemize}
Because the majority of through-going muons induce purely electromagnetic showers (Fig.~\ref{fig:Showers}), these smaller cuts are sufficient to contain the majority of the isotopes produced in the showers. Smaller-sized cuts are necessary for this subset of muons because their high frequency makes cuts on them expensive in terms of deadtime. The cuts described above cause $\sim$9\% deadtime and reduce the backgrounds produced by this subset of muons to $\sim$0.3\%.

In these calculations, we assume ideal neutron tagging with 100\% efficiency. In practice, the efficiency is slightly lower, at $\sim$98\%~\cite{JUNO:2022mxj, JUNO:2022lpc}. The small inefficiency might arise from detector electronics, which must recover within a few microseconds following high-multiplicity events in order to register delayed neutron captures, whose mean capture time is $\sim$200~$\mu$s~\cite{JUNO:2015zny}. In addition, a small fraction of neutrons may escape the scintillator volume before capture due to their longitudinal and lateral displacement relative to the parent hadronic shower, which is similar to that observed in water~\cite{Nairat:2024upg}. However, these geometric losses are modest because the neutron displacement scale is small compared to the detector size. Overall, this slightly imperfect tagging efficiency has a negligible effect on our results, as it only marginally increases the small fraction of untagged hadronic showers and changes the total remaining background by less than a few percent. Further details on the impact of neutron-tagging efficiency can be found in Ref.~\cite{Nairat:2024upg}.

For stopping muons, the dominant products are $^{12}$B and $^{8}$Li, which account for $\sim$5\% of total backgrounds above 2.3~MeV. Because of the short half-life of these isotopes, a 5~s cut around stopping muon endpoints is enough to contain all events. In terms of size, we set the radius of the cylindrical cut on stopping muons to 1.5~m. This is necessary to reduce stopping isotopes to $\sim$0.1\% due to their displacement from the muon entry point, as seen in Fig.~\ref{fig:Distances}. This causes negligible deadtime because stopping muons are rare ($\sim$4\% of all muons).

%%%%%%%%%%%%%%%%%%%%%%%%%%%%%%%%%%%%%%
\begin{figure}[t]
    \includegraphics[width=0.99\columnwidth]{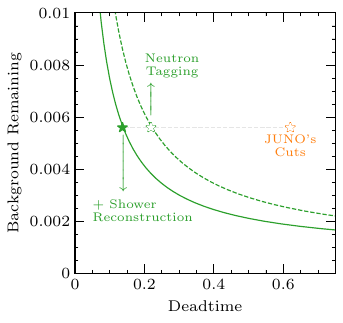}
\caption{The background rejection efficiency (fraction of remaining background) of muon cuts compared to the deadtime (fraction of signal loss) associated with these cuts. The open orange star shows the results obtained by applying JUNO's cuts~\cite{JUNO:2020hqc}. The open and filled green stars show the results of our cuts without and with shower reconstruction.}
\label{fig:Deadtime}
\end{figure}
%%%%%%%%%%%%%%%%%%%%%%%%%%%%%%%%%%%%%%

So far, we have applied cylindrical cuts uniformly along the entire muon track, which has an average length of 23.6~m. In reality, muon-induced showers are much more localized. Hadronic and electromagnetic showers of typical energies extend only 3–6~m longitudinally~\cite{Li:2015kpa, Nairat:2024upg}. Their positions, and the associated spallation products, can be inferred from the muon light profile, since shower peaks correlate with peaks in the light yield. This correlation can be exploited to restrict cylindrical cuts to the shower region, significantly reducing deadtime while maintaining the same background rejection. Such methods have already been developed for Super-K~\cite{Bays:2012wty, Super-Kamiokande:2021snn}, with Ref.~\cite{Li:2015lxa} showing how to make significant improvements.  We estimate that the deadtime from cylindrical cuts in JUNO can be reduced by at least a factor of two by limiting cuts to a $\sim$10~m region around the light-profile peak.  Recent work for KamLAND-Zen~\cite{KamLAND-Zen:2023spw} and JUNO~\cite{Zhang:2025dhu} is very encouraging, suggesting that our assumption is conservative.

Further improvements of our results can be achieved with likelihood-based methods that generalize cylindrical cuts by combining spatial and temporal information into a unified discriminator. In particular, separate likelihood functions can be constructed for muons with and without neutrons to fully exploit neutron tagging. Finally, machine learning techniques have been proven useful for spallation background reduction~\cite{Li:2018rzw, Zhang:2024ezk}. Physics-informed implementations of such methods are expected to yield even greater gains in background rejection.

%%%%%%%%%%%%%%%%%%%%%%%%%%%%%%%%%%%%%%%%%%%%%%%%%%%%%%%%%%%%%%%%%%%%%%%

\subsection{Results and comparison}

Figure~\ref{fig:Deadtime} compares the efficiency and deadtime of our cuts compared to the baseline cuts proposed in Ref.~\cite{JUNO:2020hqc}. Applying our neutron-tagging cuts reduces the spallation backgrounds above 2.3~MeV to 0.56\% of their original rate (the same value that JUNO aims for~\cite{JUNO:2020hqc}), while reducing the deadtime from 62\% to 21\%.  By using shower reconstruction and restricting the cylindrical cuts to a region of 10~m around the peak of the muon light profile, the deadtime can be further reduced to 14\%.  Thus, our techniques can reduce the deadtime by a factor of five.  This improvement in deadtime at fixed background efficiency can be traded off for a change in background efficiency at fixed deadtime or for a combination of changes.  In Fig.~\ref{fig:Deadtime}, we illustrate this by showing possible ways this could be done, obtained by varying the parameters of the cylindrical cuts.

Figure~\ref{fig:SpallationSpectra} (bottom panel) details the resulting spallation spectra after our proposed cuts. The deadtime effect on the solar neutrino signal is accounted for. The primary remaining backgrounds are $^{10}$C, $^{11}$Be, and $^8$Li. The reason that some $^{10}$C, $^{11}$Be isotopes remain is their long half-lives. $^{10}$C can be further reduced by extending the time period of the TFC cut, while $^{11}$Be can be reduced by increasing the time of the cylindrical cuts for hadronic showers. However, this would be at the cost of incurring more deadtime as demonstrated in Fig.~\ref{fig:Deadtime}. The small fraction of $^8$Li missed by our cuts is due to this isotope being primarily produced through (n,p+$\alpha$); occasionally, a shower may produce only one neutron, which subsequently interacts through this process, resulting in the production of $^8$Li without a detectable neutron capture.

\begin{figure}[t]
\includegraphics[width=0.99\columnwidth]{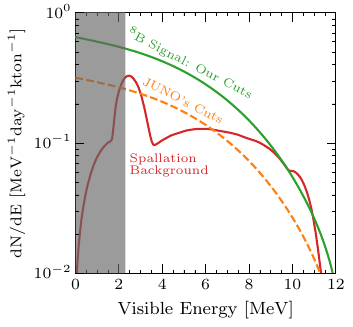}
\caption{$^8$B signal and spallation background energy spectra using our cuts and JUNO's cuts~\cite{JUNO:2020hqc}.  Our cuts have five times less deadtime.}
\label{fig:Comparison}
\end{figure}

Figure~\ref{fig:Comparison} shows the energy spectra of the $^8$B signals and total spallation backgrounds after our cuts and JUNO's proposed cuts~\cite{JUNO:2020hqc}. The signal spectrum is two times larger using our cuts due to the substantially reduced deadtime. The spallation background spectrum is similar for both, because our cuts were specifically optimized to achieve similar background rejection as JUNO's. The spallation background obtained here using JUNO's cuts is slightly larger than that in Ref.~\cite{JUNO:2020hqc} for two reasons. First, the muon rate used in their work was smaller than the one we use here (3.6~s$^{-1}$ compared to 5~s$^{-1}$), as explained previously. Second, the background spectrum in their work was obtained by using a very approximate formula to scale the yields of background isotopes as measured in Borexino and KamLAND~\cite{Borexino:2008fkj, KamLAND:2009zwo}, while we use the yields predicted by FLUKA.

These results highlight the advantage of categorizing muons by shower type --- hadronic versus electromagnetic --- and tailoring cuts accordingly. JUNO’s baseline cuts treat all muons identically, incurring a large deadtime. By contrast, our strategy achieves much better results by leveraging neutron tagging and the physical distinctions between hadronic and electromagnetic showers.

%%%%%%%%%%%%%%%%%%%%%%%%%%%%%%%%%%%%%%
%%%%%%%%%%%%%%%%%%%%%%%%%%%%%%%%%%%%%%

%%%%%%%%%%%%%%%%%%%%%%%%%%%%%%%%%%%%%%%%%%%%%%%%%%%%%%%%%%%%%%%%%%%%%%%
%%%%%%%%%%%%%%%%%%%%%%%%%%%%%%%%%%%%%%%%%%%%%%%%%%%%%%%%%%%%%%%%%%%%%%%

\section{Conclusions \& Future Work}
\label{sec:Conclusions}

JUNO has unique capabilities --- huge size, excellent energy resolution, and low backgrounds --- that make it promising for solar neutrino measurements~\cite{JUNO:2020hqc, JUNO:2022jkf, JUNO:2023zty}. However, its relatively shallow depth poses a major challenge. The high flux of cosmic-ray muons leads to abundant production of unstable isotopes through spallation processes, and the subsequent beta decays of these isotopes mimic solar neutrino events. Furthermore, directional reconstruction of the recoil electrons is not possible, which eliminates the ability to reduce backgrounds using directional correlation with the Sun.

In this paper, we present the most detailed theoretical study of muon spallation in JUNO. Expanding on prior work for scintillator detectors~\cite{Galbiati:2004wx, Galbiati:2005ft, Borexino:2011cjz, Borexino:2013cke, KamLAND:2009zwo}, we identify all relevant background isotopes, their yields, and production processes. We show that almost all isotopes are produced in hadronic showers, which are rare.  Most significantly, we show that these hadronic showers also produce neutrons, which JUNO can use to tag hadronic showers.  With our new cut techniques, JUNO could reduce its projected deadtime from 62\% to 14\%, \textit{an improvement of a factor of about 4.5}.  Further improvements should be possible.  In future work, direct measurements of the spallation and neutron yields in JUNO, along with dedicated simulations by the JUNO collaboration, will allow our results to be optimized.

\begin{figure}[t]
    \includegraphics[width=0.99\columnwidth]{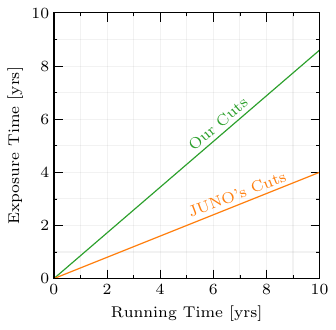}
\caption{UNO's exposure time as a function of running time in years with our cuts compared to JUNO's cuts~\cite{JUNO:2020hqc}.  Using our cuts will halve the running time required to meet a given exposure-time target.}
\label{fig:Exposure}
\end{figure}

Figure~\ref{fig:Exposure} shows that \textit{our results would allow JUNO to halve the running time required to reach the same exposure time}.  This will substantially accelerate progress toward precision solar neutrino measurements. As examples, this will improve searches for the spectrum upturn, the day-night effect, and the $hep$ flux. In the companion paper~\cite{Nairat:2025qxd}, we demonstrate that those measurements may help confirm the MSW framework. Conversely, a deviation from that could indicate the presence of new physics. 

%%%%%%%%%%%%%%%%%%%%%%%%%%%%%%%%%%%%%%
%%%%%%%%%%%%%%%%%%%%%%%%%%%%%%%%%%%%%%

%%%%%%%%%%%%%%%%%%%%%%%%%%%%%%%%%%%%%%%%%%%%%%%%%%%%%%%%%%%%%%%%%%
%%%%%%%%%%%%%%%%%%%%%%%%%%%%%%%%%%%%%%%%%%%%%%%%%%%%%%%%%%%%%%%%%%

\bigskip
\section*{Acknowledgements}

We thank the anonymous referees for their helpful and constructive feedback that greatly improved the manuscript. The work of O.~N. and J.~F.~B. was supported by National Science Foundation Grant No.\ PHY-2310018.

\clearpage
\bibliography{JUNO}
\end{document}